\documentclass[10pt]{article}
\usepackage[sectionbib,authoryear]{natbib} 
\usepackage{array}
\usepackage{amssymb}
\usepackage{mathrsfs}
\usepackage{mathptmx}
\usepackage{amsmath}
\usepackage{amssymb}
\usepackage{graphics}
\usepackage{indentfirst}
\usepackage{authblk}
\usepackage{amsfonts}
\usepackage{bm}
\usepackage{graphicx}

\newtheorem{thm}{Theorem}[section]

\newcommand{\la}{\lambda}

\title{Identity Tests for High Dimensional Data Using RMT}
\author[a,b]{Cheng Wang \thanks{wwcc@mail.ustc.edu.cn}}
\author[b]{Jing Yang}
\author[a]{Baiqi Miao}
\author[b]{Longbing Cao}
\affil[a]{Department of Statistics and Finance, University of Science and
Technology of China, Hefei, Anhui 230026, China}
\affil[b]{Advanced Analytics Institute, University of Technology Sydney, New
South Wales 2007, Australia}
\date{}

\begin{document}
\maketitle

\begin{abstract}
In this work, we redefined two important statistics, the CLRT test (Bai et.al., Ann. Stat.
\textbf{37} (2009) 3822-3840) and the LW test (Ledoit and Wolf, Ann. Stat.
\textbf{30} (2002) 1081-1102) on identity tests for high dimensional data using
random matrix theories. Compared with existing CLRT and LW tests, the new tests
can accommodate data which has unknown means and non-Gaussian distributions.
Simulations demonstrate that the new tests have good properties in terms of size and
power. What is more, even for Gaussian data, our new tests
perform favorably in comparison to existing tests. Finally, we find the CLRT is
more sensitive to eigenvalues less than 1 while the LW
test has more advantages in relation to detecting eigenvalues larger than 1.

\end{abstract}

\section{Introduction}
Suppose $X_1,\cdots,X_n$ are independent and
identically distributed (i.i.d.) $p$-dimensional random vectors with population
covariance matrix $\Sigma_p$ and our interest is to test
\begin{eqnarray} \label{h0}
H_0:~\Sigma_p=I_p ~vs.~H_1:~\Sigma_p \neq I_p,
\end{eqnarray}
where $I_p$ denotes the $p-$dimensional identity matrix. Note that the identity
matrix  in (\ref{h0}) can be replaced by any other positive definite matrix
$\Sigma_0$ through multiplying the data by $\Sigma_0^{-1/2}$.

In this work, we assume $y_n=p/n \to y \in (0, \infty)$. For canonical
statistical analysis where the sample size $n$ tends to infinity while the
dimension $p$ remains fixed, one can refer to \cite{And}. When $y<1$,
\cite{Bai09} proposed a correction to the classic likelihood ratio test (CLRT)
and derived the central limit theorem(CLT) using random matrix theories(RMTs). When
$y \geq 1$, CLRT is degenerate since the sample covariance is no longer
invertible with probability one and \cite{lw} gave a new statistics (LW test)
which could accommodate situations for any $y>0$. We note that the LW test has received
much attention in relevant literature including \cite{Schott06} who considered the
test for the equality of the smallest eigenvalues of $\Sigma_p$ and \cite{Birke05}
who extended the LW test to cases $y=0$ and $\infty$. There has also been a
substantial body of research motivated by the LW test such as \cite{Fisher10},
\cite{Sri05} and \cite{lin08}.

However, most of these results were derived under
Gaussian assumptions or equivalent conditions such as the fourth moment equals
three. The difficulty in relaxing Gaussian assumptions is due to the central limit
theorems for linear spectral statistics defined by eigenvalues. More details
can be found in \cite{Bai04} who built the CLT for linear spectral statistics
of large-dimensional sample covariance matrices under the assumption of fourth
moments and \cite{Pan08} who improved the results for general finite fourth moments.
In \cite{Bai04} and \cite{Pan08}, the authors proposed a
simplified version of classic sample covariance matrices where the means of the
data must be
known. Recently, \cite{Pan12} derived the CLT for linear spectral statistics
of classic large-dimensional sample covariance matrices. Some other important
results include \cite{Bai10}, \cite{clt2} and so on.

CLRT in \cite{Bai09} is only applicable to Gaussian data with known means and the LW
test in \cite{lw} can only be applied to Gaussian data. Since the two tests are too narrow 
for use in applications, in this work, we will redefine the CLRT and LW tests using classic
sample covariance matrices. The CLTs of the two new tests are derived in general
conditions which can accommodate data with unknown means and non-Gaussian
distributions. Simulations demonstrate that the proposed tests have good
properties in terms of size and power. What is more, even for Gaussian data, our new
tests perform favorably in comparison to existing tests. We also study the features
of each test. That is, compared with the LW test, the CLRT has its own advantages on detecting the
eigenvalues of $\Sigma_p$ near zero which means the CLRT is more sensitive to
eigenvalues less than 1 while the LW test has more advantages on detecting
eigenvalues larger than 1. In the existing literature, there is also some work which is not based on
sample covariance matrices such as \cite{Chen10} who proposes a test by
constructing estimators from the data directly. We also conduct simulations to
compare our proposed tests with the one in \cite{Chen10}.

The paper is structured as follows: Section 2 introduces the basic data
structure and establishes the asymptotic normality of the new CLRT and new LW
tests while Section 3 reports simulation studies. All the technical details include
proofs and the preliminary results in RMT are presented in the Appendix.

\section{Main Results}
We assume the observations $X_1,\cdots,X_n$ satisfy a multivariate model
(\cite{Bai96})
\begin{eqnarray} \label{data}
X_i=\Sigma_p^{1/2} Y_i+\mu,~for~i=1,\cdots,n,
\end{eqnarray}
where $\mu$ is a $p$-dimensional constant vector and the entries of
$\mathcal{Y}_n=(Y_{ij})_{p \times n}=(Y_1,\cdots,Y_n)$ are i.i.d. with $E
Y_{ij}=0$, $EY^2_{ij}=1$ and $EY^4_{ij}=3+\Delta$. Here we introduce two
versions of the sample covariance matrices. The classic one is defined as
\begin{eqnarray*}
S_n=\frac{1}{n-1} \sum_{k=1}^n (X_k-\bar{X}) (X_k-\bar{X})',
\end{eqnarray*}
where $\bar{X}=\frac{1}{n} \sum_{k=1}^n X_k$ and a simplified version takes the
form
\begin{eqnarray*}
B_n=\frac{1}{n} \sum_{k=1}^n (X_k-\mu) (X_k-\mu)'.
\end{eqnarray*}
We refer to \cite{Bai06} and \cite{Pan12} for the differences between $S_n$ and
$B_n$ in RMT. Then we can introduce CLRT in \cite{Bai09}
\begin{eqnarray} \label{bclrt}
\hat{L}_n=\frac{1}{p} tr(B_n)-\frac{1}{p} \log{|B_n|}-1,
\end{eqnarray}
where $tr$ denotes the trace.

When $X_i \sim N_p(0, I_p)$ (or $X_i \sim N_p(\mu, I_p)$ where $\mu$ is known),
\cite{Bai09} derived the CLT of CLRT
\begin{eqnarray*}
p(\hat{L}_n-(1+(1/y_n-1)\log{(1-y_n)}))\stackrel{D}{\rightarrow}\mathbb{N}(-\log
{(1-y)}/2, -2 y -2 \log{(1-y)}),
\end{eqnarray*}
where $\stackrel{D}{\rightarrow}$ denotes convergence in distribution and
$\mathbb{N}$ the normal distribution.

When $p$ is larger than sample size $n$, since the sample covariance matrix
$S_n$ is singular, \cite{lw} proposed the LW test which is defined as
\begin{eqnarray}\label{W0}
\hat{W}_n=\frac{1}{p} tr(S_n-I_p)^2-\frac{p}{n-1} (\frac{1}{p} tr(S_n))^2,
\end{eqnarray}
If $X_i \sim N_p(\mu, I_p)$ and under some other assumptions, \cite{lw} have
proven
\begin{eqnarray} \label{llw}
n \hat{W}_n \stackrel{D}{\rightarrow} \mathbb{N}(1,4).
\end{eqnarray}
In our work, we will redefine the CLRT as
\begin{eqnarray} \label{L}
L_n=\frac{1}{p} tr(S_n)-\frac{1}{p} \log{|S_n|}-1,
\end{eqnarray}
and the LW test as
\begin{eqnarray}\label{W}
W_n=\frac{1}{p} tr(S_n-I_p)^2-\frac{p}{n-1} (\frac{1}{p}
tr(S_n))^2-\frac{(1+\Delta)(n-2)(n-1)-2}{n(n-1)^2}.
\end{eqnarray}
In (\ref{L}) and (\ref{W}), noting that the statistics are defined by $S_n$
which is invariant under
the shift transformation $X_i=X_i +c$, we can assume $\mu=0$ in (\ref{data})
without loss of generality. By \cite{Bai04}, we know almost surely
\begin{eqnarray*}
\frac{1}{p}\log{|S_n|}-\frac{1}{p}\log{|\Sigma_p|} \stackrel{a.s.}{\rightarrow}
-1-(1/y-1)\log{(1-y)}\equiv d(y),
\end{eqnarray*}
which means $L_n+d(y)$ is an estimator of $(tr(\Sigma_p)-\log{|\Sigma_p|}-p)/p$.
Similarly, $W_n$ is an estimator of $tr(\Sigma_p-I_p)^2/p$.
Noting that
\begin{eqnarray*}
\Sigma_p=I_p \Leftrightarrow(tr(\Sigma_p)-\log{|\Sigma_p|}-p)/p=0
\Leftrightarrow tr(\Sigma_p-I_p)^2/p=0,
\end{eqnarray*}
therefore, $L_n$ and $W_n$ can act as the statistics to test (\ref{h0})
theoretically. Next, we will establish the asymptotic normalities of $L_n$ and
$W_n$.

\begin{thm} \label{lrt}
When $\Sigma_p=I_p$ and $p/y \to y \in(0,1)$,
\begin{eqnarray*}
p(L_n-(1+(1/y_n-1)\log{(1-y_n)})) \stackrel{D}{\rightarrow} \mathbb{N}(m,v),
\end{eqnarray*}
where $m=y (\Delta/2-1) -3\log(1-y)/2$ and $v=-2 y- 2\log{(1-y)}$.
\end{thm}
Compared with the CLRT in \cite{Bai09} which is only applicable to Gaussian data
with known means, our new CLRT can be applied to general data with unknown
means. Further, if the population mean is unknown, the CLRT in \cite{Bai09}
will behave poorly and the new one will still be applicable.

\begin{thm}\label{mo}
Under $H_0$ and $p/n \to y \in (0,\infty)$,
\begin{eqnarray*}
p W_n \stackrel{D}{\rightarrow} \mathbb{N}(0,4y^2).
\end{eqnarray*}
\end{thm}
When $E Y_{i j} \neq 0$, Theorem \ref{lrt} and \ref{mo} are still applicable
under new assumptions $E(Y_{ij}-E Y_{i j})^2=1$ and $E(Y_{ij}-E Y_{i
j})^4=3+\Delta$ by \cite{Pan12}. In Theorem \ref{mo}, when $X_i \sim N(\mu,I_p)$
that is $\Delta=0$, we can get
\begin{eqnarray*}
p (\hat{W}_n-\frac{(n-2)(n-1)-2}{n(n-1)^2}) \stackrel{D}{\rightarrow}
\mathbb{N}(0,4y^2)
 \end{eqnarray*}
which is in accordance with (\ref{llw}) by Slutsky's theorem. Further, direct
calculations can show that when $\Sigma=I_p$, the new LW test is the unique best
unbiased estimator of $\frac{1}{p}tr(\Sigma_p-I_p)^2$ by
Lehmann-Scheff$\acute{e}$ theorem when the LW test always has a $O(\frac{1}{n^2})$
bias. Therefore, our new LW test behaves better than the LW test for Gaussian
data especially when the sample size $n$ is small. Moreover, the new LW test can be
applied to data with general distributions while the existing LW test is only
applicable to Gaussian data.

Here we also mention the result of \cite{Chen10} (CZZ test) which is also an
unbiased estimator of $\frac{1}{p}tr(\Sigma_p-I_p)^2$ and based on
$\{X_1,\cdots,X_n\}$ directly. Compared with the CZZ test which does not depend on
$\Delta$, our new LW test has several advantages. First, for Gaussian data
where $\Delta=0$, the new LW test behaves better because it is the
unique best unbiased estimator of $\frac{1}{p}tr(\Sigma_p-I_p)^2$. Second,
the new LW test has a more simple formula. For example, to calculate the new LW test, we
only need the sample covariance matrix $S_n$ when the CZZ test consists of five
parts. Finally, for general distributions, simulations show that the new LW test has
a better size when $\Delta<0$. In addition, the fourth moment $\Delta$ is a
regular condition in the CLT of statistics based on a high-dimensional
sample covariance matrix. $\Delta$ appears, for example, in the work of
\cite{Bai04}, \cite{Pan08}, \cite{clt2}, \cite{Bai10} and \cite{Pan12}.
\section{Simulations}
We report results from simulation studies which were designed
to evaluate the performance of the proposed identity tests. Here, the ratio $y$
could be estimated by $y_n=p/n$.
To evaluate the power of the tests, two different population covariance matrices
will be considered in the simulations. We set $\Sigma^{(1)}_p = diag(1.5~I_{[0.2
p]}, I_{p-[0.2p]})$ and $\Sigma^{(2)}_p = diag(0.5~I_{[0.2 p]}, I_{p-[0.2 p]})$,
where $[x]$ denoted the
integer truncation of $x$. The diagonal covariance $\Sigma_p$ has
respectively 20\% of its diagonal elements being 1.5 or 0.5 whereas the rest
are 1.

\subsection{CLRT and New CLRT}
In this part, we will study our new CLRT and the existing CLRT. Since the
existing CLRT in \cite{Bai09} can only deal with Gaussian variables with known
means, we only consider Gaussian variables with zero means in our simulations.
Table 1 shows the empirical sizes and powers of our
redefined CLRT and the existing CLRT for Gaussian variables $Y_{ij}\sim
\mathbb{N}(0,1)$. The nominal test level is set
at 5\% and all results are based on $10^3$ replications.
\begin{table}[h]
\centering
\caption{Performances of the redefined CLRT and existing CLRT}
\begin{tabular}{ccccccc}
\hline
 &\multicolumn{3}{c}{Redefined CLRT}&\multicolumn{3}{c}{Existing CLRT}\\
$n$&$y=0.25$&$y=0.5$&$y=0.75$&$y=0.25$&$y=0.5$&$y=0.75$\\
\hline
\multicolumn{7}{c}{$\Sigma_p=I_p,~Y_{ij}\sim \mathbb{N}(0,1)$}\\
40&0.077&0.072&0.076&0.071&0.061&0.062\\
80&0.062&0.061&0.062&0.060&0.056&0.055\\
160&0.054&0.053&0.054&0.053&0.052&0.053\\
\multicolumn{7}{c}{$\Sigma_p=\Sigma^{(1)}_p,~~Y_{ij}\sim \mathbb{N}(0,1)$}\\
40&0.220&0.182&0.177&0.214&0.168&0.162\\
80&0.397&0.342&0.281&0.397&0.337&0.275\\
160&0.819&0.769&0.632&0.816&0.762&0.625\\
200&0.926&0.889&0.815&0.925&0.895&0.816\\
\multicolumn{7}{c}{$\Sigma_p=\Sigma^{(2)}_p,~~Y_{ij}\sim \mathbb{N}(0,1)$}\\
40&0.371&0.369&0.272&0.370&0.367&0.278\\
80&0.841&0.749&0.618&0.840&0.762&0.641\\
160&1&1&0.986&1&0.999&0.990\\
\hline
\multicolumn{7}{c}{$\Sigma_p=I_p,~~Y_{ij}\sim \mathbb{N}(1/4,1)$}\\
100&0.066&0.051&0.059&0.689&0.837&0.834\\
\hline
\end{tabular}
\end{table}

From Table 1, we know even for Gaussian variables with known means, our new CLRT
is comparable to one in \cite{Bai09}. As $p$ and $n$ both have increased, the
sizes or powers of the two tests are quite close to 5\% or 1 and make not much
difference. Further, for the same sample size $n$, when $y$ gets smaller,
the sizes are closer to 5\% and the powers are closer to 1. The
explanation is that if we only have $n$ samples, when $p$ gets smaller (that is
$y$ gets smaller), our redefined CLRT or existing CLRT, as the estimator of
$(tr(\Sigma_p)-\log{|\Sigma_p|}-p)/p$, will become more accurate.

If the true mean $\mu$ is not zero and we still use the CLRT in \cite{Bai09},
the result of the last part in Table 1 is the experiment for $Y_{ij}\sim
\mathbb{N}(1/4,1)$. It can be found that the existing CLRT behaves very poorly
and our CLRT is still applicable.

\subsection{LW, New LW and CZZ tests}
From the definitions of the LW and the new LW tests, we know
$n(W_n-\hat{W}_n)=O(\frac{1}{n})$ which means the two statistics are quite
similar for normal distributions. Therefore, our first experiment is to
investigate the empirical sizes of the LW, the new LW and the CZZ tests on Gaussian data
with a small sample size. Results based on $10^4$ replications are reported in
Table \ref{tab2}.
\begin{table}
\centering
\caption{Sizes of the new LW, LW and CZZ tests on Gaussian data.}
\begin{tabular}{cccccccccc}
\hline
 &\multicolumn{3}{c}{New LW test}&\multicolumn{3}{c}{LW
test}&\multicolumn{3}{c}{CZZ test}\\
$p$&$n=5$&$10$&$50$&$n=5$&$10$&$50$&$n=5$&$10$&$50$\\
\hline
5&0.100&0.086&0.067&0.106&0.088&0.067&0.163&0.112&0.073\\
10&0.107&0.086&0.068&0.115&0.087&0.068&0.167&0.098&0.069\\
50&0.108&0.082&0.059&0.114&0.083&0.059&0.158&0.095&0.063\\
100&0.114&0.085&0.057&0.122&0.086&0.057&0.157&0.100&0.060\\
\hline
\end{tabular}
\label{tab2}
\end{table}

We observe from Table \ref{tab2} that the sizes of the LW and the new LW tests are
always better than the CZZ test for normal distributions. The reason is due to the fact that
the LW and the new LW tests are based on the sample covariance matrix $S_n$
which is a completely sufficient statistic for $\Sigma_p$ for Gaussian data. When
the sample size $n$ is small such as $n=5$ in the experiments, the new LW test has
better sizes than the LW test and when $n$ is large, the performances of the LW and
the new LW tests are quite similar. This is because the new LW test is always the
unique best unbiased estimator of $\frac{1}{p}tr(\Sigma_p-I_p)^2$ while the LW test
has a $O(\frac{1}{n^2})$ bias.

For non-Gaussian data, from Theorem \ref{mo}, we know that the CLT of the LW test depends
on $\Delta$ and it is not
reasonable that \cite{Chen10} assumed $\Delta=0$ even for
gamma random vectors. From Theorem \ref{mo}, when $\Sigma_p=I_p$ and
$Y_{i,j}\sim Gamma[4,0.5]$, by Slutsky's theorem we know
\begin{eqnarray*}
(n \hat{W}_n-1)/2 \stackrel{D}{\rightarrow} \mathbb{N}(0.75,1).
\end{eqnarray*}
However, in \cite{Chen10}, the authors still thought $(n \hat{W}_n-1)/2
\stackrel{D}{\rightarrow} \mathbb{N}(0,1)$. Moreover, since
$P(Z>\Phi^{-1}(0.95))=0.185$ where $Z\sim \mathbb{N}(0.75,1)$ and $\Phi$ is the
distribution function of standard norm variables, this explains why the size of
the LW test in \cite{Chen10} is near 0.185 not 5\%.

Here we will repeat part of the simulations in \cite{Chen10} using the new LW test.
Two scenarios are considered\\
  (I) $Y_{i,j}~i.i.d.~\sim Gamma[4,0.5]$ where $\Delta=1.5$;\\
  (II) $Y_{i,j}~i.i.d.~\sim Uniform[0,2 \sqrt{3}]$ where $\Delta=-1.2$.\\
Simulation results are reported in Table \ref{tab3} where the performances are
based on $10^4$ replications. It is noted that Table 1 in \cite{Chen10} has the
results for the sphericity test, not the identity test. Since the authors claimed the
simulation results for the identity
test followed very similar patterns to those of the sphericity test, here
for comparison purposes, we will still use Table 1 in \cite{Chen10} for the identity test.
\begin{table}
\centering
\caption{Performances of the new LW test and the CZZ test on non-Gaussian
data}
\begin{tabular}{ccccccccc}
\hline
 &\multicolumn{4}{c}{LW test}&\multicolumn{4}{c}{CZZ test}\\
$p$&$n=20$&$n=40$&$y=60$&$n=80$&$n=20$&$n=40$&$n=60$&$n=80$\\
\hline
\multicolumn{9}{c}{$Gamma~random~vectors$}\\
38&0.1165&0.0923&0.0833&0.0807&0.0861&0.0767&0.0707&0.0704\\
55&0.1110&0.0828&0.0822&0.0734&0.0810&0.0697&0.0694&0.0658\\
89&0.1042&0.0840&0.0704&0.0655&0.0792&0.0685&0.0591&0.0590\\
159&0.0998&0.0795&0.0649&0.0626&0.0754&0.0653&0.0583&0.0588\\
\hline
\multicolumn{9}{c}{$Uniform~random~vectors$}\\
38&0.0574&0.0517&0.0490&0.0501&0.0678&0.0565&0.0527&0.0530\\
55&0.0614&0.0592&0.0539&0.0543&0.0728&0.0659&0.0563&0.0563\\
89&0.0548&0.0503&0.0530&0.0563&0.0650&0.0548&0.0570&0.0582\\
159&0.0592&0.0556&0.0546&0.0518&0.0718&0.0607&0.0563&0.0535\\
\hline
\end{tabular}
\label{tab3}
\end{table}

\begin{figure}
\includegraphics[scale=0.75]{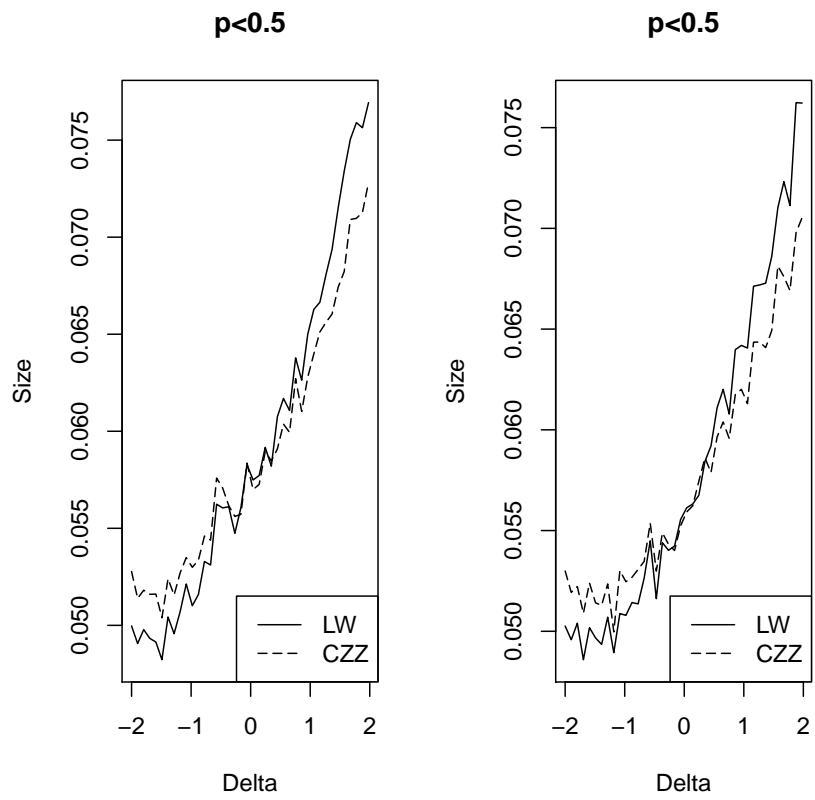}
\caption{Realized sizes of the LW test and the CZZ test for different fourth moment
$3+\Delta$.}
\label{fig1}
\end{figure}

From Table \ref{tab3}, we can see that the new LW test is not as bad as shown in
Table 1 of \cite{Chen10}.
The sizes of the new LW test and the CZZ test are comparable and when $p$ and $n$
increase, the sizes both tend to the nominal 5\% level. Specially, for Gamma
data ($\Delta=1.5$), the CZZ test has a better size while the sizes of the new LW test
are closer to the nominal level for Uniform variables ($\Delta=-1.2$). From
Table \ref{tab3}, it seems like that the LW test has a better size when $\Delta<0$
compared with the CZZ test. To verify this point, we designed another experiment to
investigate the differences of the sizes between
the new LW test and the CZZ test. For the new simulations, we set
\begin{eqnarray*}
 P(Y_{ij}=-\sqrt{\frac{1-\gamma}{\gamma}})=1-P(Y_{ij}=\sqrt{\frac{\gamma}{1-\gamma}})=\gamma \in (0,1),
\end{eqnarray*}
then it is easy to show
\begin{eqnarray*}
 E
Y_{ij}=0,~EY^2_{ij}=1,~\Delta=EY^4_{ij}-3=\frac{(1-\gamma)^2}{\gamma}+\frac{\gamma^2}{1-\gamma}-3.
\end{eqnarray*}
In the experiment, by adjusting $\gamma$ in $(0,0.5)$ or $(0.5,1)$, we can get the
results for different $\Delta$.
The results based on $10^5$ replications are reported in Figure \ref{fig1} where
$p=50,~n=100$.

We observe from Figure \ref{fig1} that when $\Delta<0$, the new LW test has a better
size and the CZZ test is better for $\Delta>0$ which is consistent with results
in Table \ref{tab3}. Here we can see the performances of the new LW test and the CZZ
test are similar being around $\Delta=0$ which is a little different from the results
for Gaussian data ($\Delta=0$). Another interesting result is that when $\Delta$
increases, the sizes of the new LW test and the CZZ test become worse although the CZZ test
does not depend on $\Delta$. We hope these questions can be addressed in future
studies.
\subsection{The powers of the new LW and CLRT tests}
\begin{figure}
\includegraphics[scale=0.750]{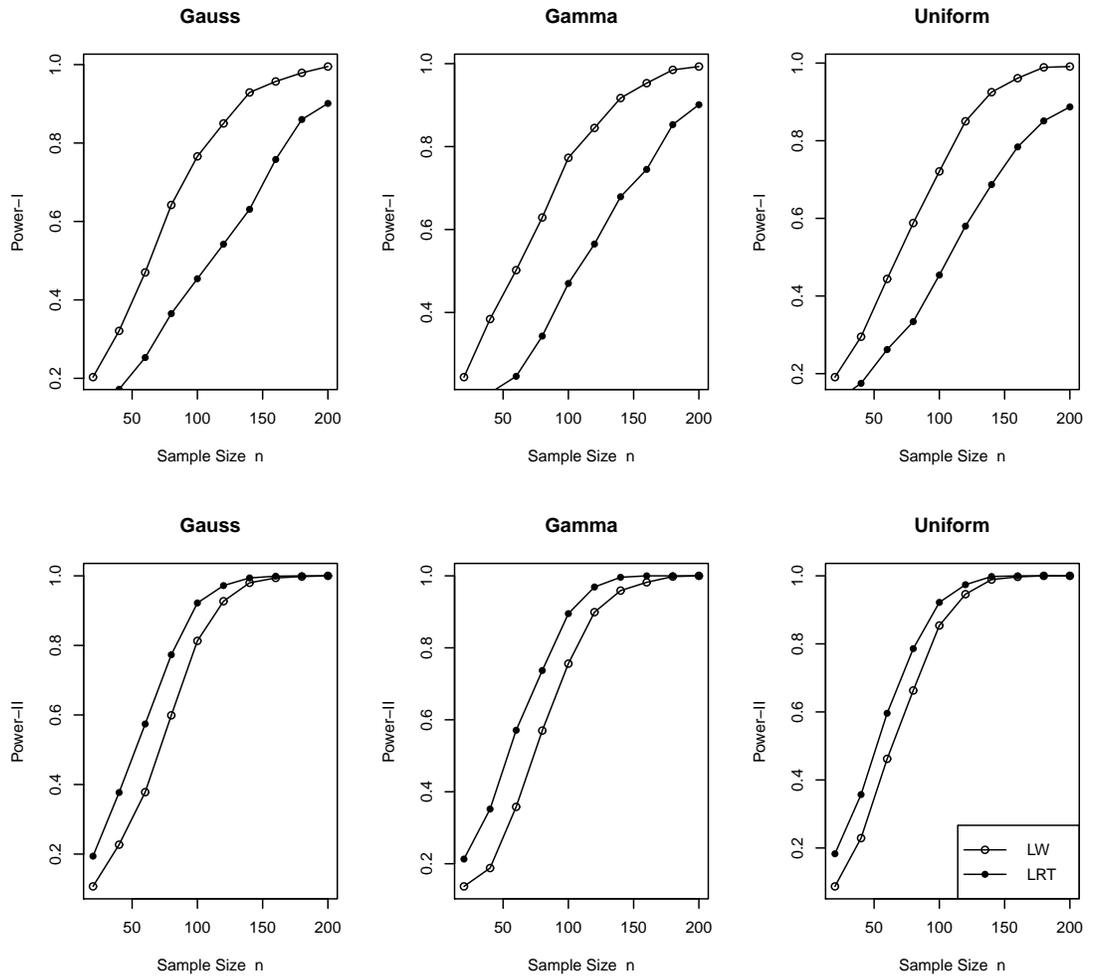}
\caption{Realized powers of the LW test and the CLRT for different sample size $n$ and
$p/n=0.5$. The top three are for $\Sigma_p=\Sigma_p^{(1)}$ and the bottom three
are for $\Sigma_p=\Sigma_p^{(2)}$.}
\label{fig2}
\end{figure}

Results based on $10^3$ replications are reported in Figure \ref{fig2},
and these correspond with $\Sigma_p=\Sigma_p^{(1)}$ or  $\Sigma_p=\Sigma_p^{(2)}$ for
three different distributions.
The power results in Figure \ref{fig2} show that the new CLRT and the new LW tests
both approach to 1 when $n$ is increased. For $\Sigma_p^{(1)}$, when part
of the eigenvalues of $\Sigma_p$ is larger than 1, the power of the new CLRT is
worse than the one of the new LW test and for $\Sigma_p^{(2)}$, when part of
eigenvalues is less than 1, the new CLRT behaves better than the new LW test. The
reason is due to the differences between $tr(\Sigma_p)-\log{|\Sigma_p|}-p$ and
$tr(\Sigma_p-I_p)^2$ where the former is more sensitive to small eigenvalues and
the latter has more advantages in terms of detecting large eigenvalues.
\section{Conclusions}
In this work, we modified two identity tests CLRT and LW test for high
dimensional data. Compared with the existing CLRT, the new CLRT and LW test can
accommodate data with unknown means and non-Gaussian distributions.
Even for Gaussian data, our new tests perform favorably in comparison to
existing tests.
In this paper, we also studied the features of each test which show that the
CLRT is more sensitive to eigenvalues less than 1 while the LW test has more
advantages in relation to detecting eigenvalues larger than 1.

From simulations, we found the new LW test has a better
size when $\Delta<0$ and the CZZ test is better for $\Delta>0$. The
performances of the new LW test and the CZZ
test are similar around $\Delta=0$ which is a little different compared to the results
for Gaussian data ($\Delta=0$). Another interesting result is that when $\Delta$
increases, the sizes of the new LW test and the CZZ test become worse although the CZZ test
does not depend on $\Delta$.

In addition, from the simulations, we found the CLRT is not the best one for
Gaussian variables although the CLRT came from the likelihood functions of normal
distributions. Finally, the powers of the tests (including CZZ test) depend on the
population covariance matrix. We hope these questions can be addressed in future
studies and an accurate estimator for $\Delta$ can be derived.

\section*{Acknowledgements}
The authors would like to thank the associate editor and an anonymous referee
for their helpful comments. Cheng Wang's research was supported by NSF of China Grants (No. 11101397, 71001095 and 11271347). Longbing Cao's research was supported by  Australian Research Council Discovery Grants (DP1096218 and DP1301691) and Australian Research Council Linkage Grant (LP100200774).
\section{Appendix}
\subsection{Preliminary results in RMT}
Suppose $A_n$ is an $n \times n$ Hermitian matrix with
eigenvalues $\la_1,\cdots,\la_n$. Define the empirical spectral distribution
(ESD) of $A_n$ as
$$F^{A_n}(x)=\frac{1}{n} \sum_{i=1}^n I(\la_i\leq x). $$
The limit distribution of $F^{A_n}$ is called the limiting spectral distribution
(LSD) of the sequence $\{A_n\}$.

And the Stieltjes transform of $F^{A_n}$ is given by
$$m^{F^{A_n}}(z)= \int \frac{1}{x-z} dF^{A_n}(x)= \frac{1}{n}
tr(A_n-z I_n)^{-1},$$
where $z=\mu+ i \nu \in \mathcal{C}^{+}$. By the inverse formula,
\begin{eqnarray} \label{sin}
F^{A_n}\{[a,b]\}=\lim_{v \to 0^+} \frac{1}{\pi} \int_a^b Im (m^{F^{A_n}}(x+i v))
dx.
\end{eqnarray}
Here we need another sample covariance matrix which is defined as
\begin{eqnarray}\label{SN}
\mathfrak{S}_n=\frac{1}{n} \sum_{k=1}^n (X_k-\bar{X})
(X_k-\bar{X})'=\frac{n-1}{n} S_n.
\end{eqnarray}
If the spectral norm of $\Sigma_p$ is bounded by a positive constant and
$F^{\Sigma_p}$ converges weakly to a non-random distribution $H$ as $p \to
\infty$, by \cite{Sil95} or \cite{Pan10}, with probability 1,
$F^{\mathfrak{S}_n}$ and $F^{B_n}$ tend to the same probability distribution
$F_{y,H}$, whose Stieltjes transform $m=m(z)~(z \in \mathcal{C}^{+})$ satisfies
\begin{eqnarray}\label{maineq}
m=\int \frac{1}{t(1-y-y z m)-z} d H(t).
\end{eqnarray}
Denoting $G_n(x)=p (F^{\mathfrak{S}_n}(x)-F_{y_n,H_n}(x))$, for any analytic
function $f$, $\int f(x) d G_n(x)$ converges weakly to a Gaussian variable $X_f$
under some assumptions on $\Sigma_p$ by \cite{Pan12}.

When $\Sigma_p=I_p$, $F_{y,H}$ is standard MP law $F_{y}$ whose density function
is
\begin{eqnarray*}
g_y(x)=\frac{1}{2 \pi y x }
\sqrt{((1+\sqrt{y})^2-x)(x-(1-\sqrt{y})^2)},~(1-\sqrt{y})^2\leq x \leq
(1+\sqrt{y})^2,
\end{eqnarray*}
and from (\ref{maineq}), we know
\begin{eqnarray*}
m=\frac{1}{1-y-y z m-z}.
\end{eqnarray*}
Writing $\underline{m}=y m-\frac{1-y}{z}$, by \cite{Pan12}, we have
\begin{eqnarray} \label{mean}
E X_f&=&-\frac{1}{2 \pi i} \int  \frac{y \underline{m}
f(-\frac{1}{\underline{m}}+\frac{y}{1+\underline{m}})}{(1+\underline{m}
)((1+\underline{m})^2-c \underline{m}^2)} d \underline{m}
-\frac{\Delta}{2 \pi i} \int  \frac{y \underline{m}
f(-\frac{1}{\underline{m}}+\frac{y}{1+\underline{m}})}{(1+\underline{m})^3}d
\underline{m} \nonumber \\
&&+\frac{y}{2 \pi i} \int
\frac{f(-\frac{1}{\underline{m}}+\frac{y}{1+\underline{m}})}{(1+\underline{m})(y
\underline{m}-1-\underline{m})}d \underline{m},
\end{eqnarray}
and
\begin{eqnarray} \label{var}
Var(X_f)&=&-\frac{1}{2 \pi^2}\int \int f(z_1)
f(z_2)\frac{1}{(\underline{m}(z_1)-\underline{m}(z_2))^2}d \underline{m}(z_1) d
\underline{m}(z_2) \nonumber \\
         &&-\frac{y \Delta}{ 4 \pi^2} (\int \frac{f(z)}{(1+\underline{m}(z))^2}
d\underline{m}(z))^2.
\end{eqnarray}
The contours in (\ref{mean}) and (\ref{var}) are both contained in the analytic
region for the function $f$ and both enclose the support of $F_{y_n,H_n}(x)$ for
large $n$. Moreover, the contours
in (\ref{var}) are disjoint.
\subsection{Proofs of Theorem \ref{lrt}}
Writing $f(x)=x-\log(x)-1$, we have
\begin{eqnarray*}
L_n= \int f(x) d F^{S_n}(x),
\end{eqnarray*}
and
\begin{eqnarray*}
&&p(L_n-\int f(x) d F_{y_n}(x))\\
&&=p(\int f(x) d F^{\mathfrak{S}_n}(x)-\int f(x) d F_{y_n}(x))+p(\int f(x) d
F^{S_n}(x)-\int f(x) d F^{\mathfrak{S}_n}(x))\\
&&=p(\int f(x) d F^{\mathfrak{S}_n}(x)-\int f(x) d
F_{y_n}(x))+(tr(S_n)-log|S_n|-tr(\mathfrak{S}_n)+\log|\mathfrak{S}_n|)\\
&&=p(\int f(x) d F^{\mathfrak{S}_n}(x)-\int f(x) d
F_{y_n}(x))+\frac{1}{n-1}tr(\mathfrak{S}_n)-p \log(\frac{n}{n-1})\\
&&=p(\int f(x) d F^{\mathfrak{S}_n}(x)-\int f(x) d F_{y_n}(x))+o(1).
\end{eqnarray*}
From \cite{Pan12}, $p(\int f(x) d F^{\mathfrak{S}_n}(x)-\int f(x) d F_{y_n}(x))$
converges weakly to the Gaussian variable $X_f$. Next we calculate the mean and
variance of $X_f$.
The following results have been given in \cite{Bai09}
\begin{eqnarray*}
\int f(x) d F_{y_n}(x)=1-\frac{y_n-1}{y_n}\log(1-y_n),
\end{eqnarray*}
\begin{eqnarray} \label{m11}
-\frac{1}{2 \pi i} \int  \frac{y \underline{m}
f(-\frac{1}{\underline{m}}+\frac{y}{1+\underline{m}})}{(1+\underline{m}
)((1+\underline{m})^2-c\underline{ m}^2)} d \underline{m}=-\frac{1}{2}\log(1-y),
\end{eqnarray}
and
\begin{eqnarray} \label{v11}
-\frac{1}{2 \pi^2}\int \int \frac{f(z_1)
f(z_2)}{(\underline{m}(z_1)-\underline{m}(z_2))^2}d \underline{m}(z_1) d
\underline{m}(z_2)=-2y-2 \log{(1-y)}.
\end{eqnarray}
Also, from \cite{Pan08}, we know
\begin{eqnarray*}
\frac{1}{2 \pi i} \oint (-\frac{1}{\underline{m}}+\frac{y}{1+\underline{m}}-1)
\frac{y \underline{m}}{(1+\underline{m})^3}d \underline{m}=0.
\end{eqnarray*}
 Therefore, to get $E X_g$, we still need to calculate
 \begin{eqnarray} \label{m12}
 &&\frac{y}{2 \pi i} \oint \log
(-\frac{1}{\underline{m}}+\frac{y}{1+\underline{m}})
\frac{\underline{m}}{(1+\underline{m})^3}d \underline{m} \nonumber \\
 &=& \frac{y}{2 \pi i} \oint \log
(-\frac{1}{\underline{m}}+\frac{y}{1+\underline{m}})
d(-\frac{1}{1+\underline{m}}+\frac{1}{2(1+\underline{m})^2}) \nonumber\\
 &=&  \frac{y}{2 \pi i} \oint
\frac{\frac{1}{\underline{m}^2}-\frac{y}{(1+\underline{m})^2}}{-\frac{1}{
\underline{m}}+\frac{y}{1+\underline{m}}}(\frac{1}{1+\underline{m}}-\frac{1}{
2(1+\underline{m})^2})d \underline{m} \nonumber\\
 &=& \frac{y}{2 \pi i} \oint (\frac{1+\underline{m}}{\underline{m} (y
\underline{m}-\underline{m}-1)}-\frac{y \underline{m}}{(\underline{m}+1)(y
\underline{m}-\underline{m}-1)})
(\frac{1}{1+\underline{m}}-\frac{1}{2(1+\underline{m})^2})d
\underline{m}\nonumber\\
 &=& \frac{y}{2 \pi i} \oint (\frac{1}{\underline{m} (y
\underline{m}-\underline{m}-1)}-\frac{1}{2 \underline{m} (\underline{m}+1)(y
\underline{m}-\underline{m}-1)})d \underline{m}\nonumber\\
 &=&\frac{y}{2},
 \end{eqnarray}
 and
 \begin{eqnarray} \label{m13}
 &&\frac{y}{2 \pi i} \int
\frac{f(-\frac{1}{\underline{m}}+\frac{y}{1+\underline{m}})}{(1+\underline{m})(y
\underline{m}-1-\underline{m})}d \underline{m} \nonumber\\
 &=& \frac{y}{2 \pi i} \int
\frac{-\frac{1}{\underline{m}}+\frac{y}{1+\underline{m}}-1-\log{(-\frac{1}{
\underline{m}}+\frac{y}{1+\underline{m}})}}{(1+\underline{m})(y
\underline{m}-1-\underline{m})}d \underline{m} \nonumber\\
 &=& -y+\frac{1}{2 \pi i} \int
\log{(-\frac{1}{\underline{m}}+\frac{y}{1+\underline{m}})}(\frac{1}{1+\underline
{m}}-\frac{1}{ \underline{m}-\frac{1}{y-1}})d \underline{m} \nonumber\\
 &=&-y-\log{(1-y)},
 \end{eqnarray}
where we used the following results
\begin{eqnarray*}
\oint \frac{1}{(\underline{m}+1)^k ( y \underline{m}-\underline{m}-1)}d
\underline{m}=0,~k=1,2,3,
\end{eqnarray*}
and one equality in \cite{Bai04}
\begin{eqnarray*}
\frac{1}{\pi i} \int
\log{(-\frac{1}{\underline{m}}+\frac{y}{1+\underline{m}})}(\frac{1}{1+\underline
{m}}-\frac{1}{ \underline{m}-\frac{1}{y-1}})d \underline{m}=-2 \log{(1-y)}.
\end{eqnarray*}
By (\ref{m11}), (\ref{m12}) and (\ref{m13}), $E X_f=y (\Delta/2-1)
-3\log(1-y)/2$.

Through a routine calculation, we have
\begin{eqnarray} \label{v12}
\frac{1}{2 \pi i} \oint
f(-\frac{1}{\underline{m}}+\frac{y}{1+\underline{m}})\frac{1}{(1+\underline{m}
)^2}d \underline{m}=0.
\end{eqnarray}
By (\ref{v11}) and (\ref{v12}), we have $Var(X_f)=-2y-2 \log{(1-y)}.$

  The proof of Theorem \ref{lrt} is complete.
\subsection{Proofs of Theorem \ref{mo}}
Noticing that $p(\frac{1}{p}tr(\mathfrak{S}_n)-1)$ satisfies CLT and
$\frac{1}{p}tr(\mathfrak{S}_n) \to 1, a.s.$, we have
\begin{eqnarray*}
p((\frac{1}{p}tr(\mathfrak{S}_n))^2-2
\frac{1}{p}tr(\mathfrak{S}_n)+1)=p([\frac{1}{p}tr(\mathfrak{S}_n)]-1)^2=o(1).
\end{eqnarray*}
By (\ref{SN}), we have
\begin{eqnarray*}
&&p
[\hat{W}_n-(\frac{1}{p}tr(\mathfrak{S}_n-I_p)^2-\frac{2}{n}tr(\mathfrak{S}
_n)+\frac{p}{n}))\\
&&=tr(\frac{n}{n-1}\mathfrak{S}_n-I_p)^2-tr(\mathfrak{S}_n-I_p)^2+(\frac{1}{n}
-\frac{n^2}{(n-1)^3})(tr(\mathfrak{S}_n))^2+o(1)\\
&&=(\frac{n^2}{(n-1)^2}-1)tr(\mathfrak{S}^2_n)-2
(\frac{n}{n-1}-1)tr(\mathfrak{S}_n)+(\frac{1}{n}-\frac{n^2}{(n-1)^3}
)(tr(\mathfrak{S}_n))^2\\
&&=-y^2+o(1),
\end{eqnarray*}
and
\begin{eqnarray*}
&&p(\frac{1}{p}tr(\mathfrak{S}_n-I_p)^2-\frac{2}{n}tr(\mathfrak{S}_n)+\frac{p}{n
})-\int ((x-1)^2-2 y x+y) d p(F^{\mathfrak{S}_n}(x)-F_{y_n}(x))\\
&&=(y-y_n)(2tr(\mathfrak{S}_n)-p)+p\int ((x-1)^2-2 y x+y) d F_{y_n}(x)\\
&&=2 (y-y_n)(tr(\mathfrak{S}_n)-p)\\
&&=o(1).
\end{eqnarray*}
Writing $g(x)=(x-1)^2-2 y x+y$, by \cite{Pan12}, $\int g(x) d
p(F^{\mathfrak{S}_n}(x)-F_{y_n}(x))$ converges weakly to gaussian variable $X_g$
and through a routine calculation
\begin{eqnarray*}
E X_g =y+y \Delta+y^2=y(1+\Delta+y).
\end{eqnarray*}
To get the variance $Var(X_g)$, we need
\begin{eqnarray*}
&&\oint\frac{g(z(m_1))}{(m_1-m_2)^2}d m_1 \\
&=& \oint\frac{1}{(m_1-m_2)^2}[(-\frac{1}{m_1}+\frac{y}{1+m_1}-1)^2+2 y
(-\frac{1}{m_1}+\frac{y}{1+m_1}+1)]d m_1\\
&=&\frac{4 \pi i y^2}{(m_2+1)^3}-\frac{4 \pi i y^2}{(m_2+1)^2}=-\frac{4 \pi i
y^2 m_2}{(m_2+1)^3},
\end{eqnarray*}

\begin{eqnarray*}
-\frac{2 y^2}{\pi i}\oint g(z(m))\frac{m}{(m+1)^3}d m= 4 y^2,
\end{eqnarray*}
and
\begin{eqnarray*}
\frac{1}{2 \pi i} \int g(-\frac{1}{m}+\frac{y}{1+m})\frac{1}{(1+m)^2}d m=0.
\end{eqnarray*}
Then
\begin{eqnarray*}
Var(X_g)=4 y^2.
\end{eqnarray*}
Above all, we can get
\begin{eqnarray*}
p \hat{W}_n \stackrel{D}{\rightarrow} \mathbb{N}(y,4y^2).
\end{eqnarray*}
By Slutsky's theorem, the proof of Theorem \ref{mo} is complete.

\bibliographystyle{natbib}
\bibliography{cite}
\end{document}